\documentclass[11pt,runningheads]{llncs}

\usepackage{amsmath,amsfonts,times,bbm,amscd,latexsym,amssymb,epsfig}
\usepackage[matrix,arrow,frame]{xy}

\makeatletter
\@addtoreset{corollary}{theorem}

\makeatother

\spnewtheorem*{measInterpol}{Measurement Pattern Interpolation}{\bf}{\it}
\spnewtheorem*{genMeasInterpol}{Generic Measurement Pattern Interpolation}{\bf}{\it}
\spnewtheorem*{Example}{Example}{\bf}{\rm}

\newcounter{romanum}
\renewcommand\theromanum{\textmd{\textup{(\textit{\roman{romanum}}\kern0.1ex)}}}
\newenvironment{romanum}
          {\setcounter{romanum}{0}
           \begin{list}{\theromanum}
          {\usecounter{romanum}
           \setlength{\parsep}{0pt}
				\settowidth{\labelwidth}{\textmd{\textup{(\textit{viii})}}}
           \setlength{\itemsep}{2pt}}}{\end{list}}

\renewcommand\ge		{\geqslant}      						
\renewcommand\le		{\leqslant}      						
\renewcommand\preceq {\preccurlyeq}							
\renewcommand\succeq {\succcurlyeq}							
\renewcommand\subset	{\subseteq}								
\newcommand\x			{\times}									
\newcommand\ox			{\otimes}              				
\newcommand\Z			{\mathbb Z}								
\newcommand\R			{\mathbb R}								
\newcommand\C			{\mathbb C\:\!}    					
\newcommand\e			{\mathrm e}   						   
\newcommand\SU			{\mathsf{SU}}
\newcommand\mG			{\mathcal G}

\newcommand\minus{\text{--}}

\makeatletter
\newcommand\diag{\mathop{\operator@font diag}\nolimits}
\newcommand\Lower{\mathop{\operator@font lower}\nolimits}
\newcommand\odd{\mathop{\operator@font odd}\nolimits}
\newcommand\img{\mathop{\operator@font img}\nolimits}
\newcommand\polylog{\mathop{\operator@font polylog}\nolimits}
\makeatother

\newcommand\union		{\,\cup\,}  							
\newcommand\inter		{\,\cap\,} 								
\renewcommand\setminus	{\smallsetminus}
\renewcommand\epsilon	{\varepsilon}					 	

\newcommand\paren[1]		{\left( #1 \right)}		     						 	
\newcommand\sqparen[1]	{\left[ #1 \right]}		   					  		
\newcommand\ens[1]		{\left\{#1\right\}}										
\newcommand\ket[1]		{\left\lvert#1\right\rangle\mspace{-1.5mu}}						
\newcommand\bra[1]		{\mspace{-1.5mu}\left\langle#1\right\rvert}						
\newcommand\card[1]		{\left\lvert #1 \right\rvert}							

\newcommand\comp{^{\textsf{c}}}	
\newcommand\trans{^\top}			
\newcommand\inv{^{-1}}				
\newcommand\intrans{^{\,-\!\top}}		
\newcommand\sperp{^\bot}			
\newcommand\s{\textsf{s}_}			

\renewcommand\vec[1]	{\mathbf{#1}}		
\newcommand\herm		{^\dagger}			

\renewcommand\qed{$\Box$}

\newcommand\PF			{\noindent \textbf{Proof.}~~}				
\newcommand\PFof[1]	{\noindent \textbf{Proof of #1.}~~}		
\def\endPF				{\hfill \qed \newline}						

\def\puttext(#1,#2)[#3]#4{\put(#1,#2){\put(-0.5,0){\makebox(0,0)[#3]{#4}}}}

\newcommand\ie{\textit{i.e.}}		
\newcommand\eg{\textit{e.g.}}		

\newcommand\ctrl{\text{\raisebox{0.3ex}{\small $\wedge$}}\mspace{-1.5mu}}	
\newcommand\cZ{\ctrl Z}																		

\newcommand\idop{\mathbbm{1}}

\begin{document}

\title{Quadratic Form Expansions for Unitaries}  
\author{Niel de Beaudrap\inst{1}, Vincent Danos\inst{2}, Elham Kashefi\inst{3}, Martin Roetteler\inst{4}}
\institute{	IQC, University of Waterloo \and
				School of Informatics, University of Edinburgh \and
				Laboratoire d'Informatique de Grenoble \and
				NEC Laboratories America, Inc.}

\maketitle


\vspace{-2.2ex}
\begin{abstract}
	We introduce techniques to analyze unitary operations in terms of \emph{quadratic form expansions}, a form similar to a sum over paths in the computational basis when the phase contributed by each path is described by a quadratic form over $\R$.
	We show how to relate such a form to an entangled resource akin to that of the one-way measurement model of quantum computing.
	Using this, we describe various conditions under which it is possible to efficiently implement a unitary operation $U$, either when provided a quadratic form expansion for $U$ as input, or by finding a quadratic form expansion for $U$ from other input data.
\end{abstract}

\section{Introduction}

In the one-way measurement model~\cite{RB01,RB02}, quantum states are transformed using single-qubit measurements on an entangled state, which is prepared from an input state by performing controlled-$Z$ operations on pairs of qubits, including the input system and ancillas prepared in the $\ket{+}$ state.
This model lends itself to ways of analyzing quantum computation which do not naturally arise in the circuit model, \eg\ with respect to depth complexity~\cite{BK07} and discrete structures underlying unitary operations~\cite{DK05c,BKMP07}. In this article, we present another result of this variety, by introducing \emph{quadratic form expansions}.

\begin{definition}
	\label{dfn:quadFormExpan}
	Let $V$ be a set of $n$ elements, and $I, O \subset V$ be (possibly intersecting) subsets.
	For a binary string $\vec{x} \in \ens{0,1}^V$, let $\vec{x}_I$ and $\vec{x}_O$ be the restriction of $\vec{x}$ to those bit-positions indexed by elements of $I$ and $O$, respectively.
	Then a \emph{quadratic form expansion} is a matrix-valued expression of the form
	\begin{align}
		\label{eqn:quadFormExpansion}
		U \;=\; \frac{1}{C} \sum_{\vec{x} \in \ens{0,1}^V} \e^{i Q(\vec{x})} \;\ket{\vec{x}_O}\bra{\vec{x}_I} ,
	\end{align}
	$U: \mathcal H_2^{\, \ox I} \to \mathcal H_2^{\, \ox O}$, where $Q$ is a real-valued quadratic form on $\vec x$,
	and $C \in \C$.
\end{definition}
Quadratic form expansions bear a formal similarity to a representation of a propagator of a quantum system in terms of a sum over paths.
For a unitary $U$ given as in \eqref{eqn:quadFormExpansion}, the outer product $\ket{\vec x_O}\bra{\vec x_I}$ essentially specifies a particular coefficient, in the row indexed by the substring $\vec{x}_I$ and the column indexed by $\vec{x}_O$: the amplitude of the transition between these standard basis states is proportional to a sum of complex units specified by $\vec{x}_I$, $\vec{x}_O$, and the auxiliary variables $v \in V \setminus (I \union O)$.

Representations of unitary transformations as sums over paths is a well-developed subject in theoretical physics (see e.g.~\cite{FH65,Sch81}); and a representation of unitaries as a sum over paths was used in~\cite{DHHMNO04} to provide a simple proof of $\mathsf{BQP} \subset \mathsf{PP}$.\footnote{%
Unitaries were expressed in~\cite{DHHMNO04} in terms of paths whose phase contributions are described by cubic polynomials over $\Z_2$; comments made in Section~VI of that paper essentially anticipate quadratic form expansions with discretized coefficients.
We describe how their techniques provide a means of constructing quadratic form expansions from circuits in Appendix~\ref{apx:sumPaths}.}
However, there are also examples of quadratic form expansions which arise without explicitly seeking to represent unitaries in terms of path integrals: the quantum Fourier Transform over $\Z_{2^n}$ can readily be expressed in such a form, and quadratic form expansions for Clifford group operations are implicit in the work of Dehaene and de Moor~\cite{DM03}, as we will describe in Section~\ref{sec:synthClifford}.

Given such an expression for a unitary $U$, we show how to obtain a decomposition of $U$ in terms of operations similar to those used in the one-way measurement model.
Using this connection, we demonstrate techniques involving quadratic form expansions to efficiently implement a unitary operator, when the coefficients of the quadratic form satisfies certain constraints related to ``generalized flows'' (or \emph{gflows})~\cite{BKMP07} or Clifford group operations.
In particular, we exhibit an $O(n^3 / \log n)$ algorithm to obtain a reduced \emph{measurement pattern} (an algorithm in the one-way model) for Clifford group operations from a description of how they transform the Pauli group, based on the results of~\cite{DM03}.

\section{Connection to the one-way model}
\label{sec:phaseMapDecomp}

\subsection{Review of the one-way model}

We can formulate the one-way measurement model as a way of transforming quantum states in the following way.
Given a state $\ket{\psi}$ on a set of qubits $I$ (the \emph{input system}), we embed $I$ in a larger system $V$, where the qubits of $V \setminus I$ are prepared in the $\ket{+} \propto \ket{0} + \ket{1}$ state.
We then perform entangling operations on the qubits of $V$, by performing controlled-$Z$ (denoted $\cZ$) operations on some sets of pairs of qubits.
(These operations are symmetric and commute with each other, and so we may characterize the entangling stage by a simple graph $G$ whose vertices are the qubits of $V$: we call this the \emph{entanglement graph} of the procedure.)
We then measure each of the qubits of $V$ in some sequence, except for some set of qubits $O \subset V$ (the \emph{output subsystem}) which will support a final quantum state.
We may represent the measurement result for each qubit $v$ by a bit $\s v \in \ens{0,1}$ which indexes the orthonormal basis states of the measurement.
The measurement basis for each qubit may depend on the results of previous measurements, but without loss of generality may be expressed in terms of a ``default'' basis which is used when all preceding measurements yield the result $0$.
Depending on the measurement results, a final Pauli operator may be applied to the qubits in the output subsystem $O$.\footnote{%
The reason for using the same variables $V$, $I$, and $O$ for these sets of (labels for) qubits as for the sets in Definition~\ref{dfn:quadFormExpan} will become apparent in the next section.}

In the original formulation of the one-way measurement model, the measurement bases were described by some axis of the Bloch sphere lying on the \textsf{XY} plane, which is sufficient for universal quantum computation.
It is also easy to prove that restricting this to states which are an angle $\theta \in \frac{\pi}{4} \Z$ from the \textsf{X} axis is sufficient for approximately universal quantum computation~\cite{DKP05}.
While it is reasonable to extend beyond this for choices of measurement bases~\cite{BB06}, we will only need to consider measurement bases from the \textsf{XY} plane.

\subsection{Phase map decompositions from quadratic form expansions}

Consider a unitary $U$ given by a quadratic form expansion as in~\eqref{eqn:quadFormExpansion}, where the quadratic form $Q$ is given by
\begin{align}
	\label{eqn:quadForm}
		Q(\vec x)
	\;=&\;
	\sum_{\ens{u,v} \subset V} \theta_{uv} x_u x_v	\;,
\end{align}
for some angles $\ens{\theta_{uv}}_{u,v \in V}$\,, and where the sum includes terms for $u = v$.
Note that $Q(\vec x)$ can be expressed as an expectation value $\bra{\vec x} H \ket{\vec x}$\,, where $H$ is a $2$-local diagonal operator:
\begin{align}
	\label{eqn:quadFormHamiltonian}
		H
	\;\;=\;
		\sum_{\substack{\ens{u,v} \subset V \\ u \ne v}} \!\!\theta_{uv} \bigg[\ket{1}\bra{1}_u \ox \ket{1}\bra{1}_v \bigg]
		\;\;\;+\;\;\;
		\sum_{v \in V} \theta_{vv} \ket{1}\bra{1}_v	\;\;.
\end{align}
Then we may decompose $U$ as follows:
\begin{align}
	\label{eqn:phaseMapDecomp}
		U
	\;\propto
		\!\!\sum_{\vec{x} \in \ens{0,1}^V} \!\!\ket{\vec{x}_O} \bra{\vec x}\;\! \e^{i H} \ket{\vec x} \bra{\vec{x}_I}\,
=&
			\sqparen{\sum_{\vec{y} \in \ens{0,1}^V} \!\!\ket{\vec{y}_O} \bra{\vec y}}
			\e^{i H}\!\!\:
			\sqparen{\sum_{\vec{x} \in  \ens{0,1}^V} \!\!\ket{\vec x} \bra{\vec{x}_I}}
	\notag\\[1ex]\propto&\;\;\;
		R_O \; \e^{iH} \, P_I	\;,
\end{align}
where $P_I$ is a unitary embedding which introduces fresh ancillas (indexed by $v \in I\comp = V \setminus I$) initialized to the $\ket{+}$ state, and $R_O$ is a map projecting onto the $\ket{+}$ state for all qubits in $O\comp = V \setminus O$ (tracing those qubits out afterwards).

Equation~\eqref{eqn:phaseMapDecomp} is a \emph{phase map decomposition}~\cite{BDK06} for $U$: that is, it expresses $U$ in terms of a process of postselecting observables, as follows.
Decompose $H$ into terms $H_O$, $H_1$, and $H_2$, where $H_O$ consists of the $1$-local terms on the qubits of $O$, $H_1$ consists of the $1$-local term on the remaining qubits, and $H_2$ contains the remaining terms from \eqref{eqn:quadFormHamiltonian}.
We then have $U \,\propto\, R_O\; \e^{iH_O} \e^{iH_1} \e^{iH_2} P_I$.
Note that $\e^{iH_O}$ and $\e^{iH_1}$ are simply single-qubit $Z$ rotations applied to the elements of $O$ and $O\comp$ respectively, where in each case the qubits $v$ in those sets are rotated by an angle $\theta_{vv}$\,.
Then, the composite map $\tilde{R}_O = R_O \e^{iH_1}$ projects each the state of each qubit $v \in O\comp$ onto the vector $\ket{0} + \e^{-i\theta_{vv}} \ket{1}$ for each $v \in O\comp$.
We then have $U = \e^{iH_O} \tilde{R}_O\, \e^{iH_2} P_I$, which is a decomposition of $U$ into the preparation of some number of $\ket{+}$ states, followed by a diagonal unitary operator consisting of two-qubit (fractional) controlled-$Z$ operations, followed by post-selection of states on the Bloch equator for $v \in O\comp$, and (unconditionally applied) single-qubit $Z$ rotations on the remaining qubits.
If $\theta_{uv} \in \ens{0,\pi}$ for all distinct $u,v \in V$ and for $u = v \in O$, the above describes precisely the action of a measurement-based computation in which the qubits $v \in O\comp$ are measured in the eigenbases of observables of the form $M(-\theta_{vv}) = \cos(\theta_{vv}) X - \sin(\theta_{vv}) Y$, in the special case where all measurements result in the $+1$ eigenstate (which we may label with the bit $\s v = 0$).

If we are able to extend the above into a complete measurement algorithm, with defined behavior when not all measurements yield a specific outcome, we obtain a measurement-based algorithm for $U$: we discuss this problem in the next section.
Conversely, from every measurement based algorithm, we may obtain a quadratic form expansion:

\begin{theorem}
		\label{thm:quadFormExpansionsUniversal}
		Every unitary operator on $n$ qubits may be expressed by a quadratic form expansion with $\card{I} = \card{O} = n$, and where the quadratic form has coefficients $\theta_{uv} \in \ens{0,\pi}$ for all cross-terms $x_u x_v$ and $-\pi < \theta_{vv} \le \pi$ for all terms $x_v^2$\,.
		Furthermore, any unitary can be approximated to arbitrary precision by such an expansion where we further require $\theta_{vv} \in \frac{\pi}{4}\Z$.
\end{theorem}

\PF
	From~\cite{DKP04} (and using the notation of that article), the measurement pattern $X_v^{\s u} M_u^{-\alpha} E_{uv} N_v$ performs the unitary transformation $J(\alpha) = \frac{1}{\sqrt 2}\big[\begin{smallmatrix} 1 & \e^{i\alpha} \\ 1 & -\e^{i \alpha}\end{smallmatrix}\big]$ for $\alpha \in \R$, from the state space of a qubit $u$ to that of a ``fresh'' qubit $v$.
	These operations generate $\SU(2)$, and generate a group dense in $\SU(2)$ if we restrict to $\alpha \in \frac{\pi}{4} \Z$, by~\cite{DKP05}.

	For any $n$ qubit unitary $U$, there exists a measurement pattern composed of such patterns together with two-qubit controlled-$Z$ operations (which we denote $\cZ$) which implements $U$.
	Let $G$ be the entanglement graph of this pattern, and $I$ and $O$ be the qubits defining the input space and output space (respectively) of the measurement pattern.
	By~\cite{DK05c}, in this measurement pattern, the probability of every measurement resulting in the $+1$ eigenvalue (i.e. $\s v = 0$ for all $v \in O\comp$) is non-zero.
	Then, $U \,\propto\, R_O\, \e^{iH} P_I$\,, where
	\begin{align}
			H
		\;\;=\;
			\sum_{uv \in E(G)} \pi \bigg[\ket{1}\bra{1}_u \ox \ket{1}\bra{1}_v \bigg]
			\;\;\;-\;\;\;
			\sum_{v \in O\comp} \alpha_v \ket{1}\bra{1}_v	\;\;.
	\end{align} 
	By~\eqref{eqn:phaseMapDecomp}, this yields a quadratic form expansion for $U$, with
	\begin{align}
			Q(\vec x)
		\;\;=\;\;
			\sum_{uv \in E(G)} \pi x_u x_v \;\;\;-\;\;\; \sum_{v \in O\comp} \alpha_v x_v^2	\;\;.
	\end{align}
	For a quadratic form expansion approximating $U$, it is sufficient to consider measurement patterns approximating $U$
	using angles $\alpha_v \in \frac{\pi}{4} \Z$.
\endPF

\subsection{Measurement Pattern Interpolation}

As we remarked above, the connection from quadratic form expansions to phase map decompositions may allow us to obtain an implementation for $U$, provided we can determine how to adapt measurements in case the measurements for qubits $v \in O\comp$ do not all yield the result $\s v = 0$.

In a measurement pattern performing $N$ measurements, the computation may follow any of $2^N$ branches, corresponding to the different combinations of measurement results. 
Let us call the branch in which every measurement produces the result $\s v = 0$ the \emph{positive branch} of the measurement pattern.\footnote{%
This choice of terminology refers to all measurements yielding the $+1$ eigenvalues of their respective observables $M(-\theta_{vv})$.} 
Without loss of generality, we may restrict our attention to patterns where no classical feed-forward is required in the positive branch: then, 
the positive branch of a measurement pattern is characterized by the \emph{geometry} $(G,I,O)$ of the pattern (where $G$ is the entanglement graph of the measurement algorithm, and $I, O \subset V(G)$ are the sets of qubits defining the input/output space of the pattern), and the angles $\vec{a} = \ens{\alpha_v}_{v \in O\comp}$ defining the measurements to be performed.

To extend the description of the positive branch of a measurement algorithm into a \emph{complete} measurement algorithm performing a unitary is the subject of the following problem:
\begin{measInterpol}[MPI]
	For input data $(G,I,O,\vec{a})$, describing a unitary embedding $U$ as \emph{the positive branch} of a measurement pattern with geometry $(G,I,O)$ and performing measurements $\vec{a}$, determine if there a measurement pattern $\mathfrak P$ with geometry $(G,I,O)$ which performs the transformation $U$.
\end{measInterpol}
This problem is open, and seems to be difficult in general.
We may attempt to make the problem easier by considering a more restricted problem:
\begin{genMeasInterpol}[GMPI]
	For an input geometry $(G,I,O)$, determine if there exist measurement patterns $\mathfrak P(\vec{a})$ parameterized by a choice $\vec{a}$ of measurement angles, each with geometry $(G,I,O)$, such that the pattern $\mathfrak P(\vec a)$ performs a unitary embedding for all $\vec{a}$.
\end{genMeasInterpol}
GMPI addresses, in an \emph{angle-independent} manner, the subject of the structure of measurement patterns which perform unitary transformations.
A special case of the GMPI which has been solved are those geometries $(G,I,O)$ which have a ``generalized flow'' (or \emph{gflow}), which are the ``\emph{yes}'' instances of GMPI such that the patterns $\mathfrak P(\vec{a})$ yield maximally random outcomes on all of their measurements~\cite{BKMP07}.
The following is the definition of gflows in~\cite{MP07}, for measurements restricted to the \textsf{XY} plane:\footnote{%
The original definition of gflows in~\cite{BKMP07} also allows for \textsf{YZ} plane and \textsf{XZ} plane measurements, which do not play a role either in our analysis or in~\cite{MP07}.}
\begin{definition}
	\label{dfn:gflow}
	Given a geometry $(G,I,O)$ for a measurement pattern, a \emph{gflow} is a pair $(g, \preceq)$, where $g$ is a function from $O\comp$ to subsets of $I\comp$ and $\preceq$ is a partial order, such that the following conditions hold for all $u$ and $v$ in the graph $G$:
 	\begin{subequations}
	\begin{align}
			v \in g(u) &\implies u \prec v \,,
		\\
			v \in \odd(g(u)) &\implies u \preceq v \,,
		\\
			u \in \odd&(g(u)) \,,
	\end{align}
 	\end{subequations}
	where $\odd(S)$ is the set of vertices adjacent to an odd number of elements of $S$.
\end{definition}
Here, $u \preceq v$ essentially represents, for two qubits $u$ and $v$, that $v$ is measured no earlier than $u$; a gflow then specifies an ordering in which the qubits are to be measured (with the function $g$ providing a description of how to adapt later measurements).
Mhalla and Perdrix~\cite{MP07} present an algorithm which determines if a geometry has a gflow in this sense in polynomial time, which in turn yields a polynomial time solution to the GMPI for that case.
As a result, any instance of the MPI where the geometry $(G,I,O)$ has a gflow can be efficiently solved.

A different special case of the Measurement Pattern Interpolation problem which has been solved is that where the measurement angles are restricted to multiples of $\frac{\pi}{2}$ (or slightly more generally, where the measurement observables are Pauli operations).
In this case, as noted in~\cite{BB06}, no measurement adaptations are necessary, and the corrections can be determined via the stabilizer formalism~\cite{GotPhD}.

In the following section, we apply these solutions to special cases of the MPI to describe how to synthesize implementations for a unitary $U$ given by a quadratic form expansion.

\section{Synthesis via measurement pattern interpolation}
\label{sec:synthesis}

In order to apply the partial solutions to the MPI described above, it will be useful to define the following:
\begin{definition}
	\label{dfn:quadraticFormGeometry}
	For a quadratic form expansion	
	\begin{align}
		\label{eqn:quadFormExpansionAgain}
		\frac{1}{C} \sum_{\vec{x} \in \ens{0,1}^V} \e^{i Q(\vec{x})} \;\ket{\vec{x}_O}\bra{\vec{x}_I}
		&&
		\text{where}\quad
		Q(\vec x)
	\;=&\;
		\sum_{\substack{\ens{u,v} \subset V}} \theta_{uv} x_u x_v \;\,,
	\end{align}
	the \emph{geometry induced by the quadratic form} is a triple $(G,I,O)$\,, where $G$ is a \emph{weighted} graph
	with vertex-set $V$, edge-set $\ens{uv \mid u \ne v ~\text{and}~ \theta_{uv} \ne 0}$\,,
	\ and edge-weights
	$W_G(uv) = \theta_{uv}/\pi$.
\end{definition}

Because we can require $-\pi < \theta_{uv} \le \pi$ for all $u,v \in V$\,, we may without loss of generality
restrict $G$ to have edge-weights $-1 < W_G(uv) \le 1$. We will assume that this holds for the remainder of the
article, and speak of edges being either of \emph{unit weight} or \emph{fractional weight}.

In this section, we consider the problem of synthesizing an efficient implementation of unitaries $U$ in terms of the geometry induced by a quadratic form expansion for $U$ by reduction to the solved cases of the Measurement Pattern Interpolation problem discussed in the previous section.

\subsection{Measurement pattern synthesis via gflows}
\label{sec:patternsViaGflows}

Consider a geometry $(G,I,O)$ induced by a quadratic form expansion for a unitary embedding $U$, where $G$ has only edges of unit weight: then $(G,I,O)$ is also a geometry for a measurement pattern.
To obtain a measurement pattern for $U$, it suffices to find a gflow for $(G,I,O)$: in that case, by Theorem~2 of~\cite{BKMP07}, for any choice of measurement angles $\vec{a} = \ens{\alpha_v}_{v \in O\comp}$, we may consider the pattern 
\begin{align}
	\label{eqn:gflowMeasPatt}
	\sqparen{{\prod_{u \in O\comp}}^{\!\succeq\;}
		\Bigg( \bigotimes_{\substack{v \in \odd(g(u)) \\ v \ne u}} Z_v \Bigg)
		\Bigg( \bigotimes_{v \in g(u)} X_v \Bigg)
		M_u^{\alpha_u}}\!
	\sqparen{\prod_{u \sim v} E_{uv}}\!
	\sqparen{\prod_{u \in I\comp} N_u}
\end{align}
where the left-hand product may be ordered right-to-left in any linear extension of the order $\preceq$, and $\sim$ denotes the adjacency relation of $G$.
This pattern thus steers the reduced state after every measurement to the state which would occur if the result had been the $+1$ eigenvalue.
Every branch of the pattern then performs the same operation as the positive branch, and so the pattern implements a unitary operation $U$.
To obtain a pattern in standard form (with corrections only on output qubits), it is sufficient to propagate the corrections to the left, absorbing them into the measurement bases.

In~\cite{MP07}, an $O(n^4)$ algorithm is provided to determine whether or not a geometry $(G,I,O)$ has a gflow where every qubit is to be measured in the \textsf{XY} plane (and obtain one in the case that one exists), where $n = \card{V(G)}$.
The measurement pattern of \eqref{eqn:gflowMeasPatt} can be constructed in time $O(n^2)$ by first producing a pattern where corrections undo byproduct operations after each measurement, commuting these corrections to the end, and simplifying; the resulting pattern will have $O(n)$ operations each with complexity $O(n)$. Thus:
\begin{theorem}
	For a unitary embedding $U$ given as a quadratic form expansion with geometry $(G,I,O)$ with unit edge-weights, there is an $O(n^4)$ algorithm which either determines that $(G,I,O)$ has no gflow, or constructs a measurement pattern consisting of $O(n^2)$ operations\footnote{
	These operations may involve measurement angles of arbitrary precision. A corresponding approximate measurement pattern may use $O(n^2 + n \polylog(n/\epsilon))$ operations by the Solovay-Kitaev Theorem~\cite{KSV02}, where $\epsilon$ is the precision of the coefficients of $Q$.}
	implementing $U$ (using measurement angles of arbitrary precision), where $n = \card{V(G)}$.
\end{theorem}

\subsection{Circuit synthesis via flows}
\label{sec:circuitsViaFlows}

A geometry $(G,I,O)$ which has fractional edges lies, at first glance, outside of the domain of the Measurement Interpolation Problems described above.
However, given a quadratic form expansion with such a geometry, we may still be able to synthesize a circuit for a unitary $U$ represented by that expansion by considering \emph{flows}, which correspond to gflows where the function $g$ maps each vertex $v \in O\comp$ to a singleton set: we may say $(f, \preceq)$ is a flow if and only if $(g_{\scriptscriptstyle f},\preceq)$ is a gflow, where $g_{\scriptscriptstyle f}(v) = \ens{f(v)}$.

Geometries which have flows are a solvable special case of the GMPI, where the resulting measurement patterns are very ``circuit-like''.
Specifically, the positive branch of a measurement pattern whose geometry has a flow can be represented by a circuit with the following characteristics~\cite{DK05c}:
\begin{itemize}
\item
	edges of the form $v \, f(v)$ for $v \in O\comp$ correspond to $J(-\alpha_v)$ gates on some wire, separating two wire segments which we label $v$ and $f(v)$;
\item
	edges $uv \in E(G)$ for $u \ne f(v)$ and $v \ne f(u)$ correspond to $\cZ$ operations acting on the wire segments labelled by $u$ and $v$;
\item
	wires whose initial segments are labelled by vertices of $I$ accept arbitrary input states, while those labelled by vertices $I\comp \setminus \img(f)$ take input $\ket{+}$.
\end{itemize}

In the above formulation, the edges of the form $v \, f(v)$ can be interpreted as implementing single-qubit teleportation, in which case a fully entangling unitary is important in order to transfer the information of the ``source'' qubit to the ``target'' qubit upon measurement.
However, considering the analysis of~\cite{DK05c}, it is not important that the edges of the second kind above be fully entangling operations: using such edges to represent fractional powers of $\cZ$ will also yield unitary circuits.
This motivates the following definition:
\begin{definition}
	Suppose $(G,I,O)$ is a geometry of a quadratic form expansion for a unitary transformation $U$.
	We may say that $(f,\preceq)$ is a \emph{fractional-edge flow} for $(G,I,O)$ if it is a flow for that geometry, and for all $ab \in E(G)$ with $W_G(ab) < 1$, we have $f(a) \ne b$ and $f(b) \ne a$. 
\end{definition}
If $(G,I,O)$ has a fractional-edge flow, we may synthesize a circuit from a quadratic form expansion for $U$ using the description above, where edges $ab$ of fractional weight correspond to $\cZ^{W_G(ab)}$ gates on the wire segments labelled by $a$ and $b$ rather than simple $\cZ$ gates.
We will make use the following easily verified Lemma to consider how to compose/decompose quadratic form expansions:
\begin{lemma}
	\label{lemma:quadFormExpanComposition}
	Let $U_1$, $U_2$ be matrices given by quadratic form expansions
	\begin{align}
			U_j
		\;=&\;\;
			\frac{1}{C_j} \! \sum_{\vec x \in \ens{0,1}^{V_j}} \e^{i Q_j(\vec x)} \; \ket{\vec x_{O_j}} \bra{\vec x_{I_j}}	\;.
	\end{align}
	In the following, $C = C_1 C_2$\,, and sums are over $\ens{0,1}^{V_1 \union V_2}$.
	\begin{romanum}
	\item
		If $V_1 \inter V_2 = I_2 = O_1$\,, then $U_2 U_1 \,=\,
			\frac{1}{C} \sum\limits_{\vec x} \; \e^{\,i Q_1(\vec x) \,+\, i Q_2(\vec x)} \, \ket{\vec x_{O_2}} \bra{\vec x_{I_1}} .
		$\vspace{1ex}
	\item
		If $V_1$ and $V_2$ are disjoint, then $\smash{U_1 \ox U_2 \,=\,
			\frac{1}{C} \sum\limits_{\vec x} \; \e^{\, i Q_1(\vec x) \,+\, i Q_2(\vec x)} \, \ket{\vec x_{O}} \bra{\vec x_{I}} ,
		}$
		where $I = I_1 \union I_2$ and $O = O_1 \union O_2$\,.
	\end{romanum}
\end{lemma}

We prove the circuit construction given by inducting on the number of edges of fractional weight.
For the base case, if $(G,I,O)$ has no fractional-weight edges at all, we may synthesize a circuit for $U$ as above, as it corresponds to a normal measurement pattern with a flow, and so falls under the analysis of~\cite{DK05c}.
We may then induct for geometries with fractional edge-weights if we can show we can decompose the geometry into ones with fewer fractional edge-weights.

\begin{figure}[tb]
	\begin{center}
		\setlength\unitlength{0.074mm}
			\begin{picture}(1350,460)(0,0)
			\put(0,40){
				\put(20,0){\includegraphics[width=380\unitlength]{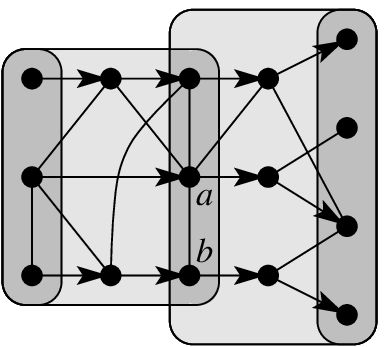}}
				\puttext(50,12)[c]{$I$}
				\puttext(365,-23)[c]{$O$}
				\puttext(140,12)[c]{$V_1$}
				\puttext(280,-23)[c]{$V_3$}
				\put(127,299){\qbezier(15,30)(40,30)(60,0)}
				\puttext(135,330)[r]{$V_2$}
				\puttext(190,390)[c]{\large $(G,I,O)$}
				}
			\puttext(480,205)[c]{\large$=$}
			\put(540,40){
				\put(20,40){\includegraphics[width=220\unitlength]{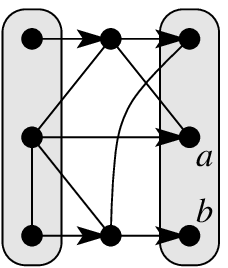}}
				\puttext(50,12)[c]{$I$}
				\puttext(215,12)[c]{$V_2$}
				\puttext(130,370)[c]{$(G_1,I,V_2)$}
			}
			\puttext(850,210)[c]{\large$\circ$}
			\put(900,40){
				\put(20,40){\includegraphics[width=65\unitlength]{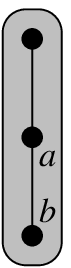}}
				\puttext(50,12)[c]{$V_2$}
				\puttext(50,370)[c]{$(G_2,V_2,V_2)$}
			}
			\puttext(1050,210)[c]{\large$\circ$}
			\put(1100,40){
				\put(20,0){\includegraphics[width=220\unitlength]{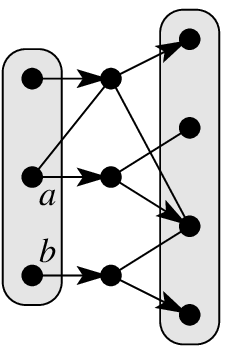}}
				\puttext(50,12)[c]{$V_2$}
				\puttext(205,-23)[c]{$O$}
				\puttext(120,370)[c]{$(G_3,V_2,O)$}
			}
			\end{picture}
		\caption{\label{fig:decompose} Illustration of the decomposion of a quadratic form expansion about an edge $ab$, expressed in terms of geometries. 
		$V_2$ is a set of maximal vertices under the constraint of being bounded from above, by the vertices $a$ and $b$, in a partial order $\preceq$ associated with a fractional-edge flow. 
		Arrows represent the action of the corresponding fractional-edge flow function, $f$. 
		}
	\end{center}
 	\vspace{-1.7em}
\end{figure}

For any arbitrary fractional edge $ab \in E(G)$ and each each $z \in O$, we may define $m(ab,z)$ to be the maximal vertex $v \in V(G)$ in the ordering $\preceq$ subject to $z$ being in the orbit of $v$ under $f$ (that is, $z = f^\ell(v)$ for some $\ell \ge 0$), such that at least one of $v \preceq a$ or $v \preceq b$ holds.
For a set $S \subset V(G)$, let $G[S]$ represent the subgraph of $G$ induced by $S$ (i.e. by deleting all vertices in $G$ not in $S$). Then, define the following subgraphs of $G$, and corresponding geometries:
\begin{itemize}
\item
	Let $V_2$ be the set of vertices $m(ab,z)$ for each $z \in O\comp$: it is easy to show that $a,b \in V_2$. Let $G_2 = G[V_2]$,
	and let $\mG_2 = (G_2, V_2, V_2)$.

\item
	Let $V_1$ be the set of vertices $u \in V(G)$ such that $u \preceq v$ for some $v \in V_2$; let $G_1 = G[V_1] \setminus \ens{uv \,\big|\, u,v \in V_2}$\,; and let $\mG_1 = (G_1, I, V_2)$.

\item
	Let $V_3$ be the set of vertices $u \in V(G)$ such that $u \succeq v$ for some $v \in V_2$; let $G_3 = G[V_3] \setminus \ens{uv \,\big|\, u,v \in V_2}$; and let $\mG_3 = (G_3, V_2, O)$.
\end{itemize}
This decomposes the geometry $(G,I,O)$ into three geometries with fractional-edge flows, as illustrated in Figure~\ref{fig:decompose}. 

Let $Q_1$ be a quadratic form on $\ens{0,1}^{V_1}$ consisting of the terms $x_u x_v$ of $Q$ for $u \in V_1$ or $v \in V_1$,
but not both; $Q_2$ be a quadratic form on $\ens{0,1}^{V_2}$ consisting of the terms $x_u x_v$ of $Q$ for \emph{distinct}
$u, v \in V_2$; and similarly let $Q_3$ be defined on $\ens{0,1}^{V_3}$, and consist of the remaining terms of $Q$.
Then $Q_1$, $Q_2$, and $Q_3$ define quadratic form expansions for some operations $U_1$, $U_2$, and $U_3$ (respectively)
with geometries $\mG_1$, $\mG_2$, and $\mG_3$ (respectively).
\begin{itemize}
\item
	$U_2$ in particular will be a product of operations $\cZ^{W_G(uv)}$ for distinct $u,v \in V_2$\,, as it is a quadratic form expansion whose input and output indices coincide. Then $U_2$ can be represented as a circuit with a wire for each $u \in V_2$, with fractional controlled-$Z$ gates $\cZ^{W_G(uv)}$ for each edge $uv \in E(G)$.
\item
	Both $\mG_1$ and $\mG_3$ have fractional-edge flows, but fewer fractional edges than $(G,I,O)$.
	By induction, $U_1$ and $U_3$ are also unitary embeddings, and have circuits with wire-segments connected by $J(\theta_v)$ gates (where $\theta_v$ are the coefficients of the terms $x_v^2$ in each quadratic form) and possibly fractional $\cZ$ gates (as in the case for $U_2$).
\item
	In the circuits described above, the terminal wire-segments for $U_1$ and (a subset of) the initial wire-segments for $U_3$ have the same labels as the wires for $U_2$\,.
	The composite circuit for $U_3 U_2 U_1$ can then use these labels to arrive at a unified labelling of its' wire-segments.
\end{itemize}
Because $Q_1(\vec{x}_{_{^{V_1}}}) + Q_2(\vec{x}_{_{^{V_2}}}) + Q_3(\vec{x}_{_{^{V_3}}}) = Q(\vec{x})$ for all $\vec{x} \in \ens{0,1}^{V}$ by construction, the composite operation $U_3 U_2 U_1$ can differ from $U$ by at most a scalar factor by Lemma~\ref{lemma:quadFormExpanComposition}; so the circuit obtained implements the operation $U$.

In~\cite{MP07}, an $O(kn)$ algorithm is provided to determine whether or not a geometry $(G,I,O)$ has a flow, and obtain one if it exists, where $n = \card{V(G)}$ and $k = \card{O}$.
For each edge $uv$, we may check whether one of $W_G(uv) = 1$ or $\big[ u \ne f(v)$ and $v \ne f(u) \big]$ holds: if all edges satisfy this constraint, the circuit described above is well-defined.
By iterating through the vertices of $V(G)$ in an arbitrary linear extension of $\preceq$\,, we may construct the circuit described above can be constructed in time $O(m)$, and the size of the resulting circuit will also be $O(m)$, where $m = \card{E(G)}$. By an extremal result~\cite{BP07}, any geometry with a flow has $m \le kn$: thus, the total running time of this algorithm is $O(kn)$.

\begin{figure}[tb]
	\begin{center}
	\frame{%
	\begin{minipage}{0.96\textwidth}
	\vspace{0.5ex}
		\setlength{\unitlength}{0.112mm}
		\begin{minipage}[b]{0.45\textwidth}
		\begin{picture}(590,220)(-10,40)
			\put(0,10){\includegraphics[width=452\unitlength]{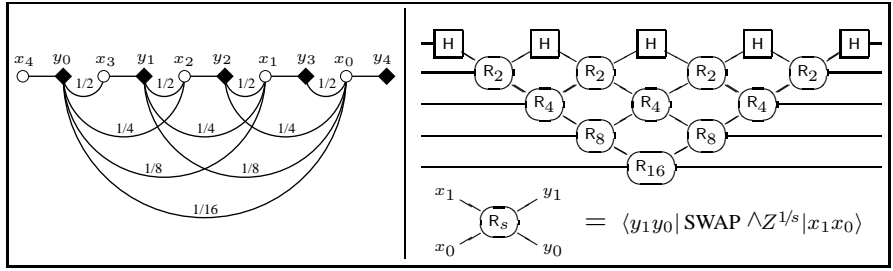}}
			\put(10,200){%
				\puttext(0,-2)[c]{$\scriptstyle x_4$}
				\puttext(45,0)[c]{$\scriptstyle y_0$}
				\puttext(97,-2)[c]{$\scriptstyle x_3$}
				\puttext(145,0)[c]{$\scriptstyle y_1$}
				\puttext(190,-2)[c]{$\scriptstyle x_2$}
				\puttext(235,0)[c]{$\scriptstyle y_2$}
				\puttext(290,-2)[c]{$\scriptstyle x_1$}
				\puttext(335,0)[c]{$\scriptstyle y_3$}
				\puttext(380,-2)[c]{$\scriptstyle x_0$}
				\puttext(425,0)[c]{$\scriptstyle y_4$}		
				}
		\end{picture}
		\end{minipage}%
		\vrule
		\begin{minipage}[b]{0.6\textwidth}
		\begin{center}
		\normalsize
		\newcommand\rud{\ar@{-}[rd]\ar@{-}[ru]}
		\newcommand\rd{\ar@{-}[rd]}
		\newcommand\ru{\ar@{-}[ru]}
		\newcommand\Rg[1]{*+[F-:<6pt>]{\scriptscriptstyle \mathsf R_{#1}}}
		\newcommand\Hg{*+[F]{\scriptscriptstyle \mathsf H}}
		\xymatrix @R=0ex @C=1ex {
			\ar@{-}[r]		&	\Hg\rd	& 					&	\Hg\rd		& 					&	\Hg\rd		&						&	\Hg\rd		&					&	\Hg	&	\ar@{-}[l]		\\
			\ar@{-}[rr]		&				& \Rg{2}\rud	&					&	\Rg{2}\rud	&					&	\Rg{2}\rud		&					&	\Rg{2}\ru	&			&	\ar@{-}[ll]		\\
			\ar@{-}[rrr]	&				&					&	\Rg{4}\rud	&					&	\Rg{4}\rud	&						&	\Rg{4}\ru	&					& 			&	\ar@{-}[lll]	\\
			\ar@{-}[rrrr]	&				&					&					&	\Rg{8}\rud	&					&	\Rg{8}\ru		&					&					&			&	\ar@{-}[llll]	\\
			\ar@{-}[rrrrr]	&				&					&					&					&	\Rg{16}\ru	&						&					&					&			&	\ar@{-}[lllll]	
			}

		\quad\xymatrix @R=0ex @C=1ex
		{ *[d]{\scriptscriptstyle x_1\;\;} \rd	& & *[d]{\scriptscriptstyle \;\;y_1} \\ & \Rg{s}\rud & & & {\!\!\!\!\!\!=\;\; \scriptstyle  \bra{y_1 y_0} \,\text{SWAP} \;\cZ^{1\!/\!s} \ket{x_1 x_0}} \\ *[u]{\scriptscriptstyle x_0\;\;} \ru & & *[u]{\scriptscriptstyle \;\;y_0} }\hfill~
		\end{center}
		\end{minipage}
	\vspace{0.5ex}
	\end{minipage}}
	\end{center}
	\vspace{-1em}
	\caption{\label{fig:qftSynth}
		The geometry for the quadratic form expansion of the QFT for $\Z_{32}$, and the corresponding circuit due to~\cite{FDH04}.
		In the geometry (on the left), input vertices are labelled by circles, output vertices by lozenges, and fractional edges are labelled with their edge-weights.}
	\vspace{-1.5em}
\end{figure}

In the case $\card{I} = \card{O}$, a flow function $f$ is unique if it exists, by~\cite{Beau06}; so in this case, if $(G,I,O)$ has a flow but there is an edge $v \, f(v)$ of fractional weight, there is no fractional-weight flow for $(G,I,O)$. We then have:

\begin{theorem}
	For a unitary transformation $U$ given as a quadratic form expansion with geometry $(G,I,O)$, there is an $O(kn)$ algorithm which either determines that $(G,I,O)$ has no fractional-edge flow, or constructs a circuit consisting of $O(kn)$ operations\footnote{%
	These operations may consist of $J(\alpha)$ gates and fractional $\cZ$ gates of arbitrary precision. A corresponding circuit using a finite elementary gate set may be of size $O(kn \polylog(kn/\epsilon))$ by the Solovay-Kitaev Theorem~\cite{KSV02}, where $\epsilon$ is the precision of the coefficients of $Q$.}
	implementing $U$, where $n = \card{V(G)}$ and $k = \card{O}$.
\end{theorem}

\begin{Example}
	The Fourier Transform over $\Z_{2^n}$ is given by the matrix formula
	\begin{align}
		\mathcal F_n
		\;\;=\;\;
			\frac{1}{\sqrt{2^n}}
			\sum_{\vec{x}, \vec{y} \in \ens{0,1}^n}
			\e\big.^{\smash{2 \pi i \Big[ \sum\limits_{h=0}^{n-1} 2^h x_h \Big] \Big[ \sum\limits_{j=0}^{n-1} 2^j y_j \Big]/2^n}} \ket{\vec y} \bra{\vec x},
	\end{align}
	which is a quadratic form expansion; its quadratic form can be given by
	\begin{align}
		Q(\vec x, \vec y) \;=\; \sum_{h = 0}^{n-1} \sum_{j = 0}^{n-1-h} \frac{2^{(h+j)}}{2^{n-1}} \,\pi x_h y_j \;.
	\end{align}
	This has a fractional-edge flow for all $n$. Figure~\ref{fig:qftSynth} illustrates this geometry for $n = 5$, and the circuit (due to~\cite{FDH04}) which may be synthesized from it.
\end{Example}

\subsection{Synthesizing measurement patterns for the Clifford group}
\label{sec:synthClifford}

If a quadratic form expansion has a geometry whose edges all have unit weight, and its' other coefficients are multiples of $\frac{\pi}{2}$, then it corresponds to the positive branch of a measurement pattern which measures only $X$ or $Y$ observables.
A measurement pattern of this sort, if it performs a unitary operation, performs a Clifford group operation in particular.

An algorithm of Aaronson and Gottesman~\cite{AG04} can produce a circuit of size $O(n^2 / \log n)$ in classical deterministic time $O(n^3 / \log n)$ for a Clifford group operation $U$ acting on $n$ qubits, from a description of how $U$ transforms Pauli operators by conjugation.
By converting the circuit into a measurement-based algorithm, and performing the graph transformations of~\cite{HEB04} to remove auxiliary qubits, we may obtain a pattern of at most $3n$ qubits\footnote{%
In~\cite{BB06}, Clifford operations on $n$ qubits are described as having minimal patterns for are described as requiring at most $2n$ qubits; however, this only holds up to local Clifford operations on the output qubits.}
in time $O(n^4 / \log n)$.
Building on the results of~\cite{DM03}, we show how to classically compute such a minimal pattern in time $O(n^3 / \log n)$ by solving the MPI for a quadratic form expansion for $U$.

\subsubsection{Obtaining a quadratic form expansion.}

For the sake of completeness, we outline the relevant results of~\cite{DM03}.
Define the following notation for bit-flip and phase-flip operators on a qubit $t$ out of a collection $\ens{1,\ldots,n}$:
\begin{align}
		P_t
	\;\;=&\;\;
		X_t	\,,
	&
		P_{n+t}
	\;\;=&\;\;
		Z_t	\,.
\end{align}
Let $\diag(M) \in \Z_2^{\, m}$ represent the vector of the diagonal elements of any square boolean matrix $M$; and let $\vec{d}(M) = \diag\left(M\trans\! \left[\begin{smallmatrix} 0 & \idop_n\! \\ 0 & 0 \end{smallmatrix} \right] M\right) \in \Z_2^{\, 2n}$
for a $2n \x 2n$ matrix $M$ over $\Z_2$. Then, we may represent an $n$ qubit unitary $U$ by a $2n \x 2n$ boolean matrix $C$ and a vector $\vec{h} \in \ens{0,1}^{2n}$, whose coefficients are jointly given by
\begin{align}
	\label{eqn:LeuvenTableau}
		U P_t U\herm
	\;\;=&\quad
		i\big.^{d_t(C)} \big(-1\big)^{h_t}
		\bigotimes_{j = 1}^n \sqparen{ Z_j^{\, C_{(n+j)t}} X_j^{\, C_{jt}} }
\end{align}
for each $1 \le t \le 2n$.~ (Note that the factor of $i^{d_t(C)}$ is only necessary to ensure that the image of $P_t$ is Hermitian, and does not serve as a constraint on the value of $C$ as a matrix.)
We will call an ordered pair $(C,\vec h)$ a \emph{Leuven tableau} for a Clifford group element $U$ if it satisfies (\ref{eqn:LeuvenTableau}).\footnote{%
Note that the block matrix $\big[\;\! C\trans \, \vec{h} \;\!\big]$ is similar to a \emph{destabilizer tableau} as defined in~\cite{AG04}.}

Provided a Leuven tableau $(C,\vec h)$ for a Clifford group operation $U$, \cite{DM03}~provides a matrix formula for $U$ which we may obtain for $U$, as follows. Decompose $C$ as a block matrix $C = \left[ \begin{smallmatrix} \,E\, & \,F\, \\ \,G\, & \,H\, \end{smallmatrix} \right]$ with $n \x n$ blocks, and then find invertible matrices $\tilde{R}_1, \tilde{R}_2$ over $\Z_2$ such that $\tilde{R}_1\inv G \tilde{R}_2 = \left[ \begin{smallmatrix} \,0\; & \!0\; \\ \,0\; & \idop_r \end{smallmatrix}\right]$ for some $r < n$~ (using \eg\ the decomposition algorithm of~\cite{PMH03} to obtain $\tilde{R}_1$ and $\tilde{R}_2$ in terms of elementary row operations).
Then, define the matrices
\begin{align}
		\left[ \begin{matrix}
			\,\tilde{E}_{11}\, & \,\tilde{E}_{12}\, \\[1ex]
			\,\tilde{E}_{21}\, & \,\tilde{E}_{22}\,
		\end{matrix} \right]
	\;=&\;\;
		\tilde{R}_1\trans E \tilde{R}_2	\,,
	&
		R_1
	\;=&\;\;
		\tilde{R}_1	\;,
	&
		R_2
	\;=&\;\;
		\left[\begin{matrix}
			\,\tilde{E}_{11}\inv\, & 0 \\[1ex]
			\;0\; & \;\idop_r\;
		\end{matrix}\right]\trans\!\!
		\tilde{R}_2\trans	\;,
\end{align}
where $\tilde{E}_{11}$ is taken to be a block of size $(n-r) \x (n-r)$. We may then obtain the block matrices
\begin{align}
		\left[\begin{smallmatrix}
			\idop_{n \minus r} & 		E_{12}		& F_{11} & F_{12} \\[1ex]
					E_{21}		& 		E_{22} 		& F_{21} & F_{22} \\[1ex]
					0				&			0			& H_{11} & H_{12}	\\[1ex]
					0				&	\idop_r	& H_{21} & H_{22}
				 \end{smallmatrix}\right]
	\;\;=&\;\;
		\left[\begin{smallmatrix} R_1\trans & 0 \\ 0 & R_1\inv \end{smallmatrix}\right]
		C
		\left[\begin{smallmatrix} R_2\trans & 0 \\ 0 & R_2\inv \end{smallmatrix}\right]	\,,
\end{align}
and use these to construct the $n \x n$ boolean matrices
\begin{align}
		M_{br}
	\;&=\;\;
		\left[\begin{matrix} \;F_{11} + E_{12} H_{21}\; & \;E_{12}\; \\[1ex] \;E_{12}\trans\; & \;E_{22}\; \end{matrix}\right]\,,
	&
		M_{bc}
	\;=&\;\;
		\left[\begin{matrix} \;0\; & \;H_{21}\trans\; \\[1ex] \;H_{21}\; & \;H_{22}\; \end{matrix}\right]\,.
\end{align}
Next, define 
\begin{align}
	\begin{split}
			\vec{d}_{br}
		&=\;
			\diag(M_{br}) \,,
		\\
			L_{br}
		\;&=\;
			\Lower\!\big(M_{br} + \vec{d}_{br} \vec{d}_{br}\trans\big) \,,
	\end{split}
	&
	\begin{split}
			\vec{d}_{bc}
		&=\;
			\diag(M_{bc}) \,,
		\\
			L_{bc}
		\;&=\;
			\Lower\!\big(M_{bc} + \vec{d}_{bc} \vec{d}_{bc}\trans\big) \,,
	\end{split}
\end{align}
where $\Lower(M)$ is the strictly lower-triangular part of a square matrix $M$ (with all other coefficients set to $0$).
Finally, define $\Pi_r = \left[\begin{smallmatrix} 0 & 0 \\ 0 & \idop_r\! \end{smallmatrix}\right]$ and $\Pi_r\sperp
= \idop_n - \Pi_r$ for the sake of brevity, and let\footnote{%
The vector formulas given here for $\vec{t}$ and $\vec{h}_{bc}$ may be obtained by repeated application of Theorem~2 of~\cite{DM03}.}
\begin{align}
		\vec{t}
	\;\;=&\;\;
		\left[\begin{matrix} \,\idop_n \,&\, 0 \;\end{matrix}\right] \vec{h}
		\;+\;	\diag\Big( \sqparen{ R_2\inv \Pi_r  } \!L_{br} \sqparen{ R_2\inv \Pi_r  }\trans \Big)		\;,
	\\[0.5ex]
	\begin{split}
		\vec{h}_{bc}
	\;\;=&\;\;
		 \left[\begin{matrix}\; 0 \,&\, R_2\intrans \,\end{matrix}\right] \vec{h}
		\;+\;	R_2\intrans \diag\Big(
					R_2\trans \Big[ L_{bc} \,+\, \Pi_r M_{bc} 
	\\[-0.5ex]&\mspace{100mu}
					\,+\, \paren{\Pi_r\sperp + \Pi_r M_{bc}} L_{br} \paren{\Pi_r\sperp +  M_{bc} \Pi_r} \Big] R_2
				\Big)	\;.
	\end{split}
\end{align}
Then Theorem~6 of~\cite{DM03} states that the unitary operation $U$ for the Clifford operation characterized by $(C, \vec{h})$ is given by the matrix formula
\begin{align}
	\label{eqn:cliffordFormula}
	\begin{split}
		U
	\,=\:\!
		\frac{1}{\sqrt{2^r}}\!\!\!
		\sum_{\substack{\vec{x}_b \in \ens{0,1}^{n \minus r} \\ \vec{x}_c, \vec{x}_r \in \ens{0,1}^r}}\!
		\bigg[\,
			(&-1)^{\paren{
									\vec{x}_{br}\trans L_{br} \vec{x}_{br} \,+\, 
									\vec{x}_r \trans \vec{x}_c \,+\,
									\vec{x}_{bc}\trans L_{bc} \vec{x}_{bc}	\,+\,
									\vec{h}_{bc}\trans \vec{x}_{bc} 
									}} \;\x
	\\[-4ex]&
			(-i)^{\paren{\vec{d}_{br}\trans \vec{x}_{br} \,+\, \vec{d}_{bc}\trans\vec{x}_{bc}}}	\;
			\ket{R_1\big. \vec{x}_{br}}\bra{R_2\inv \vec{x}_{bc} + \vec{t}}
		\,\bigg]	\,,
	\end{split}
\end{align}
where $\vec{x}_{br} = \left[\begin{smallmatrix} \vec{x}_b \\ \vec{x}_r \end{smallmatrix}\right]$ and
$\vec{x}_{bc} = \left[\begin{smallmatrix} \vec{x}_b \\ \vec{x}_c \end{smallmatrix}\right]$ are $n$ bit boolean
vectors.

The formula in~(\ref{eqn:cliffordFormula}) shows strong similarities to a quadratic form expansion.
In particular, consider disjoint sets of indices $V_b$, $V_r$, and $V_c$, with $\card{V_b} = n-r$ and $\card{V_r} = \card{V_c} = r$.
Let $V = V_b \union V_c \union V_r$, $I = V_b \union V_c$, and $O = V_b \union V_r$, and define the following notation for $\vec{x} \in \ens{0,1}^V$\,:
\begin{align}
		\vec{x}_I
	\;=&\;
		\left[\begin{matrix} \vec{x}_b \\ \vec{x}_c \end{matrix}\right]
	\;=\;
		\left[\begin{matrix} \vec{x}_{V_b} \\ \vec{x}_{V_c} \end{matrix}\right] \in \ens{0,1}^{I}	\;,
	&
		\vec{x}_O
	\;=&\;
		\left[\begin{matrix} \vec{x}_b \\ \vec{x}_r \end{matrix}\right]
	\;=\;
		\left[\begin{matrix} \vec{x}_{V_b} \\ \vec{x}_{V_r} \end{matrix}\right] \in \ens{0,1}^{O}	\;,
\end{align}
\vspace{-2em}
\begin{align}
	\begin{split}
		Q(\vec x)
	\;=&\;\;
		\pi\Big( \vec{x}_O\trans L_{br} \vec{x}_O \,+\,
							\vec{x}_O\trans \Pi_r \vec{x}_I \,+\, \vec{x}_I\trans L_{bc} \vec{x}_I \,+\,
							\vec{x}_I\trans \vec{h}_{bc} \vec{h}_{bc}\trans \vec{x}_I	\Big)
	\\&\mspace{170mu}-\;
		\frac{\pi}{2} \Big(	\vec{x}_O\trans \vec{d}_{br}\vec{d}_{br}\trans \vec{x}_O \,+\,
									\vec{x}_I\trans \vec{d}_{bc}\vec{d}_{bc}\trans \vec{x}_I \Big)	\;.
	\end{split}
\end{align}
Then, (\ref{eqn:cliffordFormula}) is equivalent to
\begin{align}
	\label{eqn:cliffordAlmostQuadFormExpan}
		U
	\;\;=\;\;
		\frac{1}{\sqrt{2^r}}
		\sum_{\vec{x} \in \ens{0,1}^V}
			\e^{i Q(\vec x)} \ket{R_1\big. \vec{x}_O}\bra{R_2\inv \vec{x}_I + \vec{t}}	\;,
\end{align}
\begin{figure}[tb]
	\begin{center}
		\setlength\unitlength{0.07mm}
		\begin{minipage}[c]{1470\unitlength}
			\begin{picture}(1470,400)(0,0)
			\put(40,0){%
				\put(0,40){\includegraphics[width=220\unitlength]{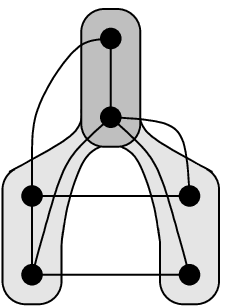}}%
				\puttext(30,20)[c]{$I$}%
				\puttext(190,20)[c]{$O$}%
				\puttext(110,370)[c]{$V_b = I \inter O$}%
			}
			\puttext(335,190)[c]{\large $\cong$}%
			\put(400,0){%
				\put(0,40){\includegraphics[width=270\unitlength]{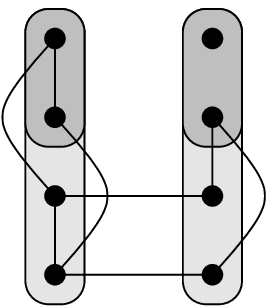}}%
				\puttext(50,20)[c]{$I$}%
				\puttext(210,20)[c]{$O'$}%
				\puttext(50,370)[c]{$V_b$}%
				\puttext(210,370)[c]{$V_{b'}$}%
			}
			\puttext(725,270)[c]{\large $\circ$}
			\put(800,0){%
				\put(0,41.5){\includegraphics[width=221\unitlength]{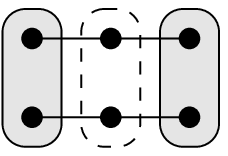}}%
				\puttext(30,370)[c]{$V_b$}%
				\puttext(110,370)[c]{$V_{a}$}%
				\puttext(190,370)[c]{$V_{b'}$}%
			}
			\puttext(1100,190)[c]{\large $=$}
			\put(1180,0){%
				\put(0,40){\includegraphics[width=270\unitlength]{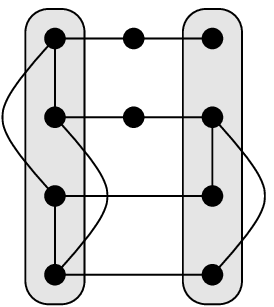}}%
				\puttext(50,20)[c]{$I$}%
				\puttext(210,20)[c]{$O'$}%
			}
			\end{picture}
		\end{minipage}
		\caption{\label{fig:idxSplit} Illustration of geometries arising from quadratic form expansions yielding the same matrix. On the left is a geometry whose inputs and output intersect; on the right is a geometry from an equivalent quadratic form expansion, constructed so that the input and output indices are disjoint.} 
	\end{center}
 	\vspace{-2.5em}
\end{figure}%
which is essentially a quadratic form expansion sandwiched between two networks of controlled-not and $X$ gates.
To obtain a simple quadratic form expansion, we would like to perform a change of variables on $\vec{x}_I$ and $\vec{x}_O$; but this cannot be done as $I$ and $O$ intersect at $V_b$, and the changes of variables do not necessarily respect the partitioning of $I$ and $O$ with respect to this intersection. However, we may add auxiliary variables in order to produce an expansion with disjoint input and output indices.
Note that
\begin{align}
		\idop_2
	\;\;=\;\;
		\sum_{\vec{x} \in \ens{0,1}^2} \delta_{x_1,x_2} \ket{x_2} \bra{x_1}
	\;\;=\;\; 
		\frac{1}{2} \sum_{\vec{x} \in \ens{0,1}^3} (-1)^{x_1 x_3 + x_2 x_3} \ket{x_2} \bra{x_1}
\end{align}
where $\delta_{x,y}$ is the Kronecker delta. Let $V_a$ and $V_{b'}$ be disjoint copies of $V_b$,
and set $V' = V \union V_a \union V_{b'}$ and $O' = V_{b'} \union V_r$. Writing $\vec{x}_a$ and $\vec{x}_{b'}$
for the restriction of $\vec{x} \in \ens{0,1}^{V'}$ to $V_a$ and $V_{b'}$, we then define
\begin{align}
		\vec{x}_I
	\;=&\;
		\left[\begin{matrix} \vec{x}_b \\ \vec{x}_c \end{matrix}\right] \in \ens{0,1}^{\, I}	\;,
	&
		\vec{x}_{O'}
	\;=&\;
		\left[\begin{matrix} \vec{x}_{b'} \\ \vec{x}_r \end{matrix}\right] \in \ens{0,1}^{\, O'}	\;,
\end{align}
\vspace{-1ex}
\begin{align}
		Q'(\vec{x}_I\,,\, \vec{x}_a\,,\, \vec{x}_{O'})
	\;=&\;\;
		\pi\Big(	\vec{x}_{O'}\trans L_{br} \vec{x}_{O'} \,+\,
							\vec{x}_{O'}\trans \Pi_r \vec{x}_I \,+\, \vec{x}_I\trans L_{bc} \vec{x}_I \,+\,
							\vec{h}_{bc}\trans \vec{x}_I	\Big)
	\notag\\&\mspace{50mu}
		\;+\;	\pi \vec{x}_I\trans \!\left[\begin{smallmatrix} \idop_{n \minus r} \\[0.5ex] 0 \end{smallmatrix}\right]\! \vec{x}_a
		\;+\;	\pi \vec{x}_{O'}\trans \!\left[\begin{smallmatrix} \idop_{n \minus r} \\[0.5ex] 0 \end{smallmatrix}\right]\! \vec{x}_a
	\notag\\&\mspace{150mu}
		\;-\;	\frac{\pi}{2} \Big(	\vec{d}_{br}\trans \vec{x}_{O'} \,+\,	\vec{d}_{bc}\trans \vec{x}_I \Big)	\;.
\end{align}
Note that the difference between the expressions for $Q'$ and $Q$ is essentially that all instances of $\vec{x}_O$ have been replaced with $\vec{x}_{O'}$ (which is independent from $\vec{x}_I$), and the presence of the terms involving $\vec{x}_a$.
(This manipulation is illustrated in Figure~\ref{fig:idxSplit} as a transformation of geometries.)
We therefore have
\begin{align}
	\label{eqn:quadFormTransform}
		\sum_{\vec{x} \in \ens{0,1}^V}
			\e^{i Q(\vec x)} &\ket{R_1\big. \vec{x}_O}\bra{R_2\inv \vec{x}_I + \vec{t}}
	\notag\\=&\;\;\;
		\sum_{\vec{x}_I,\, \vec{x}_{O'}}
			\delta_{\vec x_b, \vec x_{b'}} \, \e^{i Q'(\vec x_I,\, \vec{0},\, \vec x_{O'})} \ket{R_1\big. \vec{x}_{O'}}\bra{R_2\inv \vec{x}_I + \vec{t}}
	\notag\\=&\;\;\;	
		\frac{1}{2^{n-r}} \sum_{\vec{x} \in \ens{0,1}^{V'}}
			\e^{i Q'(\vec{x}_I,\, \vec{x}_a,\, \vec{x}_{O'})} \ket{R_1\big. \vec{x}_{O'}}\bra{R_2\inv \vec{x}_I + \vec{t}} \,.
\end{align}
Substituting the final expression of (\ref{eqn:quadFormTransform}) into (\ref{eqn:cliffordAlmostQuadFormExpan}) and performing the appropriate change of variables, we have
\begin{align}
	\label{eqn:cliffordQuadFormExpan}
		U
	\;\;=\;\;
		\frac{\sqrt{2^r}}{2^n}
		\sum_{\vec{x} \in \ens{0,1}^{V'}}
			\e^{i Q'(R_2(\vec{x}_I + \vec{t}),\, \vec{x}_a,\, R_1\inv \vec{x}_{O'})} \ket{\vec{x}_{O'}}\bra{\vec{x}_I}	\;.
\end{align}

Note that the quadratic form of the expansion in (\ref{eqn:cliffordQuadFormExpan}) has only angles $\theta_{uv}$ which are multiples of $\frac{\pi}{2}$, with $\theta_{uv} \in \ens{0, \pi}$ for $u \ne v$.
This then represents the positive branch of a one-way measurement pattern on the geometry $(G', I, O')$ of the quadratic form expansion of~\ref{eqn:cliffordQuadFormExpan}, using only $X$ or $Y$ basis measurements, and having only $n-r$ auxiliary vertices.

\subsubsection{Interpolating the measurement pattern.}

We can augment this to a measurement pattern by applying the techniques of the stabilizer formalism~\cite{GotPhD} to the stabilizer code generated by the operators $K(v) = X_v \prod_{v \sim w} Z_w$ for $v \in I\comp$ (where again $\sim$ is the adjacency relation of $G$), as follows.
To obtain the final correction, we do classical pre-processing simulating the evolution of the \emph{state space} when we perform one measurement at a time.
For each measured qubit $u$, there is an associated correction $\sigma_v$ which we may perform immediately after the measurement if we obtain the result $\s u = 1$.
We store for each qubit $v$ two boolean formulas $\beta_v$ and $\gamma_v$, representing the $X$ and $Z$ components of the accumulated corrections to be performed on $v$.
When $v$ is measured, the pending $X$ corrections will affect the result of any $Y$ measurement, and the pending $Z$ corrections will affect the result of any $X$ or $Y$ measurement, in each case by exchanging the significance of the two measurement outcomes.\footnote{%
This can be described in terms of \emph{signal shifting}, as described in~\cite{DKP04}.}
Just prior to the (simulated) measurement of $\delta_v$, let $\delta_v = \gamma_v$ if $v$ is to be measured with an $X$ observable, and $\delta_v = \beta_v + \gamma_v$ if $v$ is to be measured with a $Y$ observable.
Thus, upon measuring $v$, the following operations are accumulated into the corrections which must be performed:
\begin{itemize}
\item
	For every qubit $w$ where $\sigma_v$ acts with an $X$ or $Y$ operation, we must add $\s v + \delta_v$ to $\beta_w$;

\item
	For every qubit $w$ where $\sigma_v$ acts with a $Y$ or $Z$ operation, we must add $\s v + \delta_v$ to $\gamma_w$.
\end{itemize}
This accounts for the accumulated corrections due to the measurement of $v$ and every preceding measurement which affects it.
By simulating measurement for all of the qubits in $O\comp$ in this way, we obtain boolean formulae for the corrections on $O$ in terms of the results of the measurements: the correction to be performed for some $w \in O$ is $X^{\beta_w} Z^{\gamma_w}$, for $\beta_w$ and $\gamma_w$ constructed after all of the (simulated) measurements.
To obtain $\beta_w$ and $\gamma_w$ for all $w \in O$ in this way takes time $O(n^2)$.

It is easy to show that the resulting measurement pattern is irreducible by the techniques of~\cite{HEB04}, by the following argument.
Let $A$ denote the set of auxiliary vertices corresponding to the bit positions of $\vec{x}_a$: note that in the measurement pattern, these are all to be measured with the observable $X$, and are adjacent only to the input/output variables $\vec{x}_I$ and $\vec{x}_O$.\footnote{%
There are no square terms $x_v^2$ for $v \in A$ or cross-term $x_u x_v$ for $u,v \in A$ before the change of variables in~\eqref{eqn:quadFormTransform}), and the change of variables itself does not introduce any.}
To eliminate a vertex $v \in A$ using the methods of~\cite{HEB04} on the geometry induced by the quadratic form expansion, we would have to identify an output variable $b_0 \in O$ adjacent to $x$, and apply the graph transformation in~\cite[Proposition~1]{HEB04}.
This would result in a geometry where $b_0$ has the former neighbors of $v$ in $G$ (and in particular is not adjacent to any more removable vertices), and where a local Clifford (which is not a Pauli operator) must be applied to $b_0$ after the entangling procedure.
Because $b_0$ is not adjacent to any other auxiliary qubit after this transformation, the local Clifford cannot be undone or made into a Pauli operator by e.g. another vertex removal; then, except by extending the computational model to allow for corrections which are local Clifford operations, performing the local Clifford can only be done by introducing an auxiliary qubit (or rather, a new output qubit following $b_0$, making the latter an auxiliary qubit).
Thus:

\begin{theorem}
	For an $n$-qubit Clifford group operation $U$ given in the form of a Leuven tableau, there is an $O(n^3 / \log n)$ algorithm which produces a minimal one-way measurement pattern for $U$.
\end{theorem}

The ability to obtain a quadratic form expansion representing a reduced measurement pattern yields a more efficient algorithm to find totally reduced Clifford patterns than from using existing techniques to obtain one via the circuit model.
The quadratic form of (\ref{eqn:cliffordQuadFormExpan}) can be found from a Leuven tableau $(C, \vec h)$ in time $O(n^3 / \log n)$, which is dominated by the time required to compute $R_1$ and $R_2$.
To contrast, an approximately optimal quantum circuit for a Clifford group operation (\ie\ consisting of $O(n^2 / \log n)$ gates) can be found from a Leuven tableau in time $O(n^3 / \log n)$ by transforming it into a destabilizer tableau, and then applying the algorithm of~\cite{AG04}.
To obtain a measurement pattern from such a circuit by composing the patterns for each gate, removing vertices opportunistically (with each removal taking time $O(n^2)$), requires time $O(n^4 / \log n)$.
Thus, making use of quadratic form expansions provides us with a faster algorithm to obtain reduced measurement patterns for Clifford group operations.

\section{Conclusions and Open Problems}

We have introduced quadratic form expansions, and developed techniques which suggest that they may be useful for synthesizing efficient implementations for unitary operations.
We described conditions under which implementations may be efficiently found for unitaries specified by quadratic form expansions;
and we showed how quadratic form expansions leads to more efficient algorithms for obtaining reduced patterns for Clifford operations in the one way measurement model.

In the introduction, we mentioned that quadratic form expansions are similar in form to a sum-over-paths representation of unitary operations, which is a well-developed subject in theoretical physics.
This raises the question of whether the techniques developed here are useful e.g. for developing algorithms to simulate physical systems.
It is not known whether the solved cases of the Measurement Pattern Interpolation problem correspond to \emph{natural} (in the more literal sense) unitaries expressed as sums over paths: this question, and how to extend the solved cases of the MPI to include propagators for interesting physical systems, remain open.


\appendix
\section{Quadratic form expansions as sums over paths}
\label{apx:sumPaths}

Let $(G,I,O)$ be the geometry of a quadratic form expansion, as defined on page~\pageref{dfn:quadraticFormGeometry}.
In the special case when $(G,I,O)$ has a fractional-edge flow as defined in Section~\ref{sec:circuitsViaFlows}, the quadratic form expansion corresponds exactly to a sum over paths as described in~\cite{DHHMNO04}, for the elementary gate set of $H$, $Z^t$, and $\cZ^t$, where $t \in R$ (\ie\ admitting arbitrary $Z$ rotations and fractional controlled-$Z$ gates).
In order to demonstrate the sense in which quadratic form expansions are sums over paths in this case, and because it represents a reasonably simple algorithm for converting quantum circuits into quadratic form expansions, we now present an alternate proof of Theorem~\ref{thm:quadFormExpansionsUniversal} based on the techniques of~\cite{DHHMNO04}.
That any quadratic form expansion with geometry with a fractional-edge flow can be constructed in this way follows by reversing the construction below.

\vspace{2ex}\PFof{Theorem~\ref{thm:quadFormExpansionsUniversal}}
	Consider a quantum circuit implementing $U$ exactly, using the operations $H$, $\cZ^t$, and $Z^t$.
	Enumerate the wires of the circuit from $1$ to $k$, and for each wire $1 \le j \le k$, introduce a \emph{path label} $x_j$ for the input end of the wire, corresponding to an input bit $x_j \in \ens{0,1}$.
	We set $I = \ens{1,\ldots,k}$.
	Divide each wire into \emph{segments}, bounded on each end by either a Hadamard gate, the input terminal of the wire, or the output terminal.
	We label the wire segments with path variables: for the segments at the inputs, we apply the labels $x_j$ for $j \in I$, and we introduce new path variables to label the remaining wire segments.
	Computational paths in the circuit are then described by setting all of the the path variables $x_1 \cdots x_n$ collectively to some particular binary string in $\ens{0,1}^n$.
	The phase contribution of each paths, governing how they interfere to produce an output state for any given input state, is described by a function $\varphi(\vec x)$ depending the gates of the circuit as follows:
	\begin{romanum}
	\item
		For every Hadamard gate on a single wire, with a path variable $x_h$ labelling the segment preceding the Hadamard
		and a	path variable $x_j$ labelling the segment following the Hadamard, we add a term $x_h x_j$\,.

	\item
		For every $\cZ^t$ operation between two wires, with a path variable $x_h$ labelling the segment of one wire and $x_j$
		labelling the segment of the other wire in which the $\cZ^t$ operation is performed, we add a term $t x_h x_j$\,.

	\item
		For every $Z^t$ operation on a wire segment labelled with a path variable $x_j$\,, we add a term $t x_j^2$.
		(Because the path variable $x_j$ ranges over $\ens{0,1}$, the extra power of $2$ has no effect.)
	\end{romanum}
	In particular, the function $\varphi(\vec{x})$ is a quadratic form, where without loss of generality the coefficients may be constrained to $-1 < t \le 1$. The phase of a given path, described by a bit-string $\vec{x} \in \ens{0,1}^n$, is then given by $(-1)^{\varphi(\vec x)} = \e^{i\pi\varphi(\vec x)}$.
	Each path also has an associated amplitude of $2^{-r/2}$, where $r = n - k$ is the number of Hadamard gates in the circuit.\footnote{%
	Although it is quite reasonable to consider $\varphi$ to be simply a polynomial over $\R$, in terms of the descriptions used in Section~VI of~\cite{DHHMNO04}, one may consider $\varphi$ to be a polynomial over the ring $\R / 2\Z$.
	If we restrict to $t \in \frac{\pi}{4} \Z$, we may simplify this to the finite ring $\Z_8$ by multiplying all of the coefficients by $4$, and using it to describe powers of $\sqrt{i}$ rather than of $-1$.}

	Let $O$ be the set of indices $j$ such that some wire is labelled by the path-variable $x_j$ at its' output end.
	Then, the initial points of computational paths are described by bit-vectors $\vec{a} \in \ens{0,1}^I$\,, and the
	terminal points of paths are described by $\vec{b} \in \ens{0,1}^O$\,. The coefficients
	$U_{\vec b, \vec a}$ can then be given as the sum of the contributions of all paths beginning at $\vec{x}_I = \vec{a}$ and ending
	at $\vec{x}_O = \vec b$\,:
	\begin{align}
			U_{\vec b, \vec a}
		\;\;=&\;\;
			\frac{1}{\sqrt {2^r}}
			\sum_{\substack{\vec{x} \in \ens{0,1}^n \\ \vec{x}_I = \vec{a} \\ \vec{x}_O = \vec{b}}} \e^{i \pi \varphi(\vec x)}	\;,
	\end{align}
	which is an expression of the coefficients of $U$ as a quadratic form expansion. 

	To obtain a proof of Theorem~\ref{thm:quadFormExpansionsUniversal}, it is sufficient to note that without loss of generality we may restrict ourselves to using $\cZ^t$ gates only for $t = 1$ to implement $U$ exactly; and that to implement $U$ to arbitrary precision, it suffices to use $Z^t$ gates where $t$ is restricted to multiples of $\frac{1}{4}$.
\endPF


\begin{thebibliography}{000}

\bibitem{RB01}
R.~Raussendorf and H.~Briegel.
\newblock \emph{A one-way quantum computer}.
\newblock Physical Review Letters 86 (5188), 2001.

\bibitem{RB02}
R.~Raussendorf and H.~Briegel.
\newblock \emph{Computational model underlying the one-way quantum computer}.
\newblock Quantum Information \& Computation, vol~2 \#6 (443) 2002.

\bibitem{BK07}
A.~Broadbent and E.~Kashefi.
\newblock \emph{Parallelizing quantum circuits}.
\newblock arXiv:0704.1736, 2007.

\bibitem{FH65}
R.~P.~Feynmann, A.~R.~Hibbs.
\newblock	\emph{Quantum Mechanics and Path Integrals}.
\newblock	McGraw-Hill, New York, 1965.

\bibitem{Sch81}
L.~S.~Schulman.
\newblock	\emph{Techniques and Application of Path Integration}.
\newblock	Wiley-Interscience, New York, 1981.

\bibitem{DK05c}
V.~Danos and E.~Kashefi.
\newblock \emph{Determinism in the one-way model}.
\newblock Physical Review A 74 (052310), 2006.
\newblock arXiv:quant-ph/0506062.

\bibitem{BB06}
D.~E.~Browne and H.~J.~Briegel.
\newblock	\emph{One-way Quantum Computation --- a tutorial introduction}.
\newblock	arXiv:quant-ph/0603226 (2006).

\bibitem{BKMP07}
D.~E.~Browne, E.~Kashefi, M.~Mhalla, and S.~Perdrix.
\newblock \emph{Generalized flow and determinism in measurement-based quantum computation}.
\newblock New J. Physics vol.~9 (250), 2007.
\newblock arXiv:quant-ph/0702212

\bibitem{DHHMNO04}
C.~M. Dawson, H.~L. Haselgrove, A.~P. Hines, D.~Mortimer, M.~A. Nielsen, and T.~J. Osborne.
\newblock \emph{Quantum computing and polynomial equations over \protect{$\mathbb Z_2$}}.
\newblock Quantum Information \& Computation vol~5 \#2 (102), 2004.
\newblock arXiv:quant-ph/0408129

\bibitem{BDK06}
N.~de~Beaudrap, V.~Danos, and E.~Kashefi.
\newblock \emph{Phase map decompositions for unitaries}.
\newblock arXiv:quant-ph/0603266, 2006.

\bibitem{DKP04}
V.~Danos, E.~Kashefi, and P.~Panangaden.
\newblock \emph{The measurement calculus.}
\newblock J. ACM vol.~54, 8 (2007).
\newblock arXiv:quant-ph/0412135

\bibitem{DKP05}
V.~Danos, E.~Kashefi, and P.~Panangaden.
\newblock \emph{Parsimonious and robust realizations of unitary maps in the one-way model}.
\newblock Physical Review A vol~72 (064301), 2005.
\newblock arXiv:quant-ph/0411071

\bibitem{MP07}
M.~Mhalla and S.~Perdrix.
\newblock Finding optimal flows efficiently.
\newblock arXiv:0709.2670, 2007.

\bibitem{KSV02}
A.~Kitaev, A.~Shen, and M.~Vylalyi.
\newblock \emph{Classical and quantum computation}.
\newblock Graduate Texts in Mathematics, vol~47, American Mathematical Society, Providence RI, 2002.

\bibitem{BP07}
N.~de~Beaudrap and M.~Pei.
\newblock 		\emph{An extremal result for geometries in the one-way measurement model}.
\newblock		To appear in Quantum Information and Computation, vol. 8 \#5 (430), 2008.
\newblock		arXiv:quant-ph/0702229

\bibitem{GotPhD}
D.~Gottesman.
\newblock {\em Stabilizer codes and quantum error correction}.
\newblock PhD thesis, Caltech, 1997.
\newblock arXiv:quant-ph/9705052.

\bibitem{DM03}
J.~Dehaene and B.~De Moor.
\newblock \emph{Clifford group, stabilizer states, and linear and quadratic operations over \protect{GF(2)}}.
\newblock Physical Review A vol~68 (042318), 2003.
\newblock arXiv:quant-ph/0304125

\bibitem{HEB04}
M.~Hein, J.~Eisert, and H.~J. Briegel.
\newblock \emph{Multi-party entanglement in graph states.}
\newblock Physical Review A vol~69 (62311), 2004.
\newblock arXiv:quant-ph/0307130

\bibitem{AG04}
S.~Aaronson and D.~Gottesman.
\newblock \emph{Improved simulation of stabilizer circuits}.
\newblock Physical Review A vol~70 (052328), 2004.
\newblock arXiv:quant-ph/0406196

\bibitem{Beau06}
N.~de~Beaudrap.
\newblock Finding flows in the one-way measurement model.
\newblock arXiv:quant-ph/0611284, 2006.

\bibitem{FDH04}
A.~Fowler, S.~Devitt, and L.~Hollenberg.
\newblock \emph{Implementation of Shor's algorithm on a linear nearest neighbor qubit array}.
\newblock Quantum Information \& Computation vol~4 \#4 (237), 2004.

\bibitem{PMH03}
K.~N. Patel, I.~L. Markov, and J.~P Hayes.
\newblock Efficient synthesis of linear reversible circuits.
\newblock To appear in Quantum Information \& Computation vol~8, 2008.
\newblock arXiv:quant-ph/0302002

\end{thebibliography}
\end{document}